# Metamaterial enhancement of metal-halide perovskite luminescence

Giorgio Adamo,[1,*] Harish Krishnamoorthy,[1] Daniele Cortecchia,[2,3] Bhumika Chaudhary,[2,3] Venkatram Nalla[1], Nikolay Zheludev[1,4] and Cesare Soci[1,*]

[1] Centre for Disruptive Photonic Technologies, TPI, SPMS, Nanyang Technological University, 21 Nanyang Link, Singapore 637371
[2] Energy Research Institute @ NTU (ERI@N), Research Techno Plaza, Nanyang Technological University, 50 Nanyang Drive, Singapore 6375533
[3] Interdisciplinary Graduate School, Nanyang Technological University, Singapore 639798.
[4] Optoelectronics Research Centre & Centre for Photonic Metamaterials, University of Southampton, Southampton SO17 1BJ, UK

[*]Corresponding authors: g.adamo@ntu.edu.sg, csoci@ntu.edu.sg

**Metal-halide perovskites are rapidly emerging as solution-processable optical materials for light emitting applications. Here we adopt a plasmonic metamaterial approach to enhance photoluminescence emission and extraction of methylammonium lead iodide (MAPbI$_3$) thin films, based on the Purcell effect. We show that hybridization of the active metal-halide film with resonant nanoscale sized slits carved into a gold film can yield more than one order of magnitude enhancement of luminescence intensity, and nearly threefold reduction of luminescence lifetime. This shows the effectiveness of resonant nanostructures in controlling metal-halide perovskite light emission properties over a tunable spectral range, a viable approach toward highly efficient perovskite light emitting devices and single-photon emitters.**

Efficient nanoscale light sources are key elements for the development of advanced nanophotonic circuits and integrated thin-film optoelectronic devices. Thanks to their exceptional optoelectronic properties, hybrid organic-inorganic perovskites are gaining prominence for solid-state lighting and displays.[1,2,3,4] Limiting factors to the adoption of perovskites in photonic applications are stability[5] and low luminescence yield (low efficiency and relatively slow rate of spontaneous emission), which so far hindered the realization of perovskite-based high-speed light-emitting devices.[6,7] A number of approaches have been adopted to overcome the luminescence yield limitation, ranging from micro-/nano-structuring of the perovskite materials[8,9,10,11,12,13] to distributed feedback and microring lasing cavities.[14,15,16,17,18,19] Thanks to their ability of trapping, confining and enhancing the optical fields in nanoscale volumes, plasmonic and dielectric metamaterials have proven to be an excellent platform for the enhancement of both linear and non-linear properties of media,[14,20] and have been used with great success for multifold enhancement of luminescence in variety of active materials like quantum dots, quantum wells and dyes and ultimately enabling both lasing and spasing.[21,22,23,24] Here we exploit the design flexibility of plasmonic metamaterials to alter the luminescence of archetypical $MAPbI_3$ perovskite films hybridized with planar resonant nanoslits tailored to match their emission peak around 765 nm. We control the degree of photoluminescence enhancement by shifting the metamaterial resonance across the $MAPbI_3$ emission line and obtain more than one order of magnitude increase in light radiation from the perovskite film – correspondingly we observe a significant narrowing of the luminescence linewidth and almost three-fold reduction of the lifetime. These findings prove that hybridization of metal-halide perovskites with metamaterials could lead to the realization of more efficient and faster nanoscale integrated light emitting devices and lasers, as well as to the improvement of intrinsic radiative properties of perovskite compounds through the control of light-matter interactions at the nanoscale.

We selected a basic metamaterial design, a periodic array of nanoscale slits carved onto a continuous metallic film, which supports dipolar resonances in the visible spectral range whose wavelength is determined by the slit length. Six metamaterial nanoslit arrays (30 x 30 $\mu m^2$ squares) were fabricated by Focused Ion Beam (FIB) milling of a 30 nm thin gold layer, evaporated over a glass substrate. The arrays feature an increasing slit length $L$, ranging from 100 nm to 150 nm, in steps of 10 nm, a constant slit width W ≈ 35 nm, and a square lattice periodicity of 2$L$. Figure 1a-c show secondary electron images of three arrays with slits length $L$ = 100 nm (1a), 120 nm (1b) and 140 nm (1c), captured over a ~750 x 750 $nm^2$ field of view: they are presented side by side to produce a distinct perception of the increase in length and periodicity.

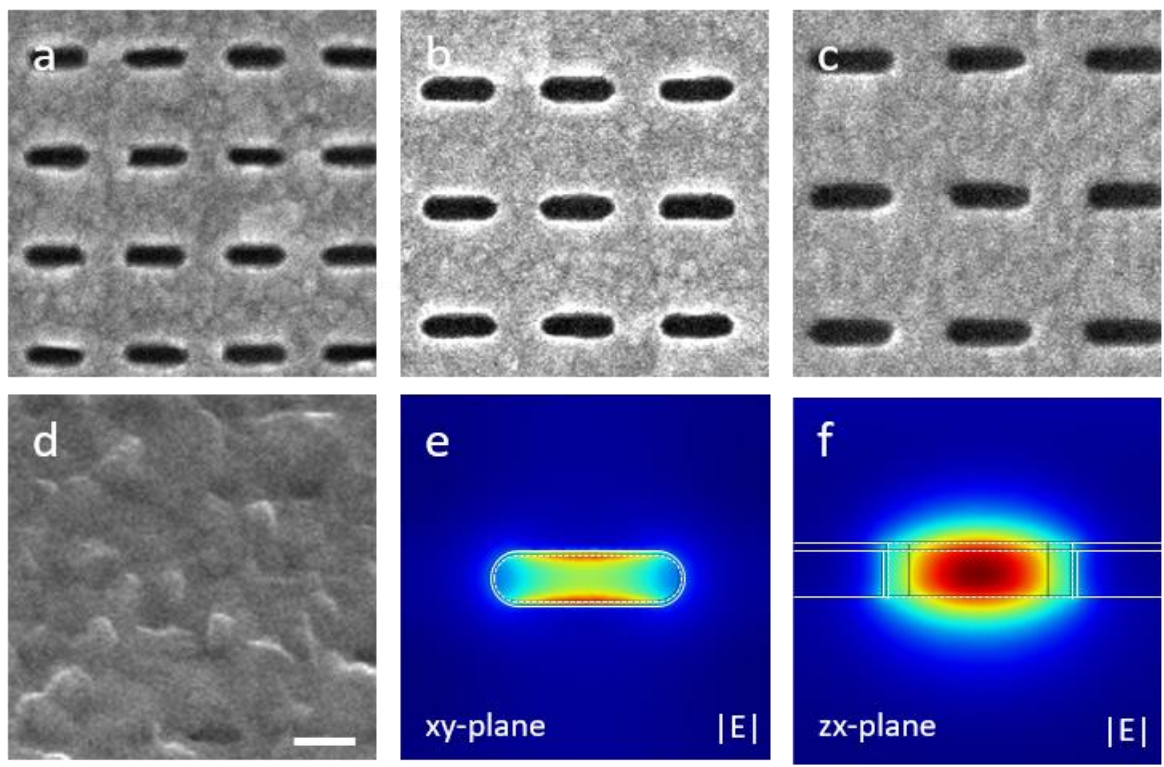

**Figure 1**: a-c) Secondary electrons images of slit metamaterials with slit length of, respectively, 100, 120 and 140 nm, carved by Focused Ion Beam (FIB) milling of a 30nm Au + 5nm $SiO_2$ film on glass substrate; d) secondary electrons image of a 150 nm slit metamaterial coated with 65nm thick $MAPbI_3$ perovskite film; scale bar is 100 nm e-f) top-view and cross-section maps of the electric field maps at resonance for a 120 nm long slit illuminated by 760nm light polarized along y, showing local field enhancement inside and in the vicinity of the slit.

The $MAPbI_3$ precursor solution was spin-coated over the entire nanostructured sample, resulting in a ~65 nm thin film, covering both the metamaterials (Figure 1d), the flat gold and a nearby patch exposing the glass substrate, to ensure as uniform as possible coating conditions for the arrays and the reference areas. To avoid direct contact between gold and the perovskite film, thus preventing luminescence quenching and the possibility of gold diffusion into the perovskite,[25] a ~5 nm $SiO_2$ layer was thermally evaporated onto the sample before spin-coating the MAPbI$_3$ solution. Three-dimensional full-wave electromagnetic simulations confirm the dipolar nature of the resonant field trapped inside the nanoslits and the enhancement of the field intensity and local density of states (LDOS), as shown in Figure 1e and 1f, respectively.

The photoluminescence spectrum of the MAPbI$_3$ perovskite film peaks around the optical bandgap energy of ~1.62 eV, with relatively narrow full-width-half-maximum (FWHM) of ~50 nm. Figures 2a and 2b show the simulated and experimentally measured optical absorption spectra of the different nanoslit metamaterials arrays fabricated, for normally incident light with polarization perpendicular to the slits (TM), before spin-coating the perovskite film. The bare metamaterial spectra feature well pronounced absorption resonances at progressively longer wavelengths with increasing slit length and crossing over the MAPbI$_3$ photoluminescence spectrum at 765 nm (Figure 2b).

The experimental optical spectra were measured under normal incidence illumination using a microphotospectrometer with a circular collection aperture of 25 μm diameter, which allowed removing edge effects from the metamaterial arrays. The simulated spectra were obtained by three-dimensional full-wave electromagnetic simulations using standard literature data for gold[26] and glass[27]. It should be noted that, in this spectral interval, optical absorption of gold is negligible, therefore the optical fields are mostly trapped within the slits or within the air/glass in its vicinity (Figures 1e-f), thus allowing strong interaction with the overlying perovskite film.

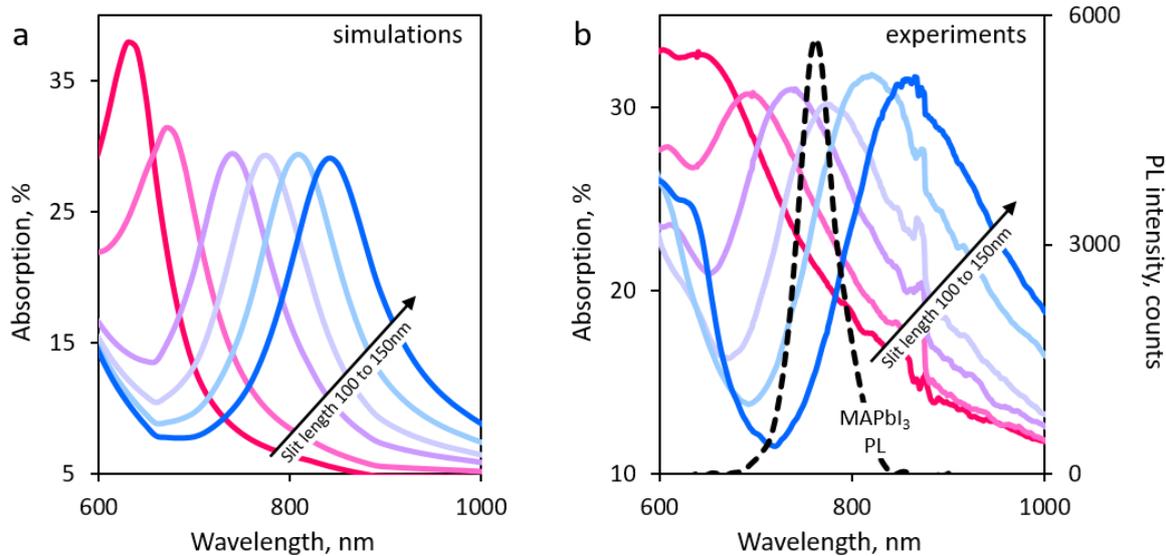

**Figure 2**: a) Simulated optical absorption resonance spectra for slit metamaterials with length varying from 100 nm to 150 nm, in 10 nm steps, (magenta to blue lines); b) experimental optical absorption resonance spectra for slit metamaterials with length varying from 100 nm to 150 nm, in 10 nm steps, overlaid with photoluminescence emission spectrum of a 65 nm thick MAPbI$_3$ perovskite film on glass (dashed black line)

The effect of the nanoslit gold metamaterials on the light emission properties of MAPbI$_3$ perovskite films was evaluated by measuring both steady-state and transient photoluminescence in three configurations: MAPbI$_3$ film on glass, MAPbI$_3$ film on gold (with ~5 nm SiO$_2$ spacer) and MAPbI$_3$ film on each of the six nanoslits metamaterial arrays (with ~5 nm SiO$_2$ spacer). In the steady-state measurements, we excited the perovskite film using a ps pulsed diode laser and measured the emitted light with a scanning monochromator spectrometer, while time resolved measurements were performed using a ~100 fs optical parametric oscillator (OPA) as pump and a streak-camera to record the photoluminescence. We have chosen an excitation wavelength of lambda=405 nm, detuned from the metamaterials' resonance, and focused the laser beam to a spot size of about 15 µm to avoid any direct effect of the pump on the interaction between MAPbI$_3$ emission and nanoslits's resonances.

The steady-state results indicate that the photoluminescence of the MAPbI$_3$ perovskite film is strongly enhanced by the interaction with the metamaterials and that the degree of enhancement can be controlled by the design of the nanoslits resonators. The measurements were performed selecting the polarization of the emitted light, whereas the polarization of the pump light was left unchanged since both the gold metamaterial and the perovskite film are polarization independent at this wavelength.

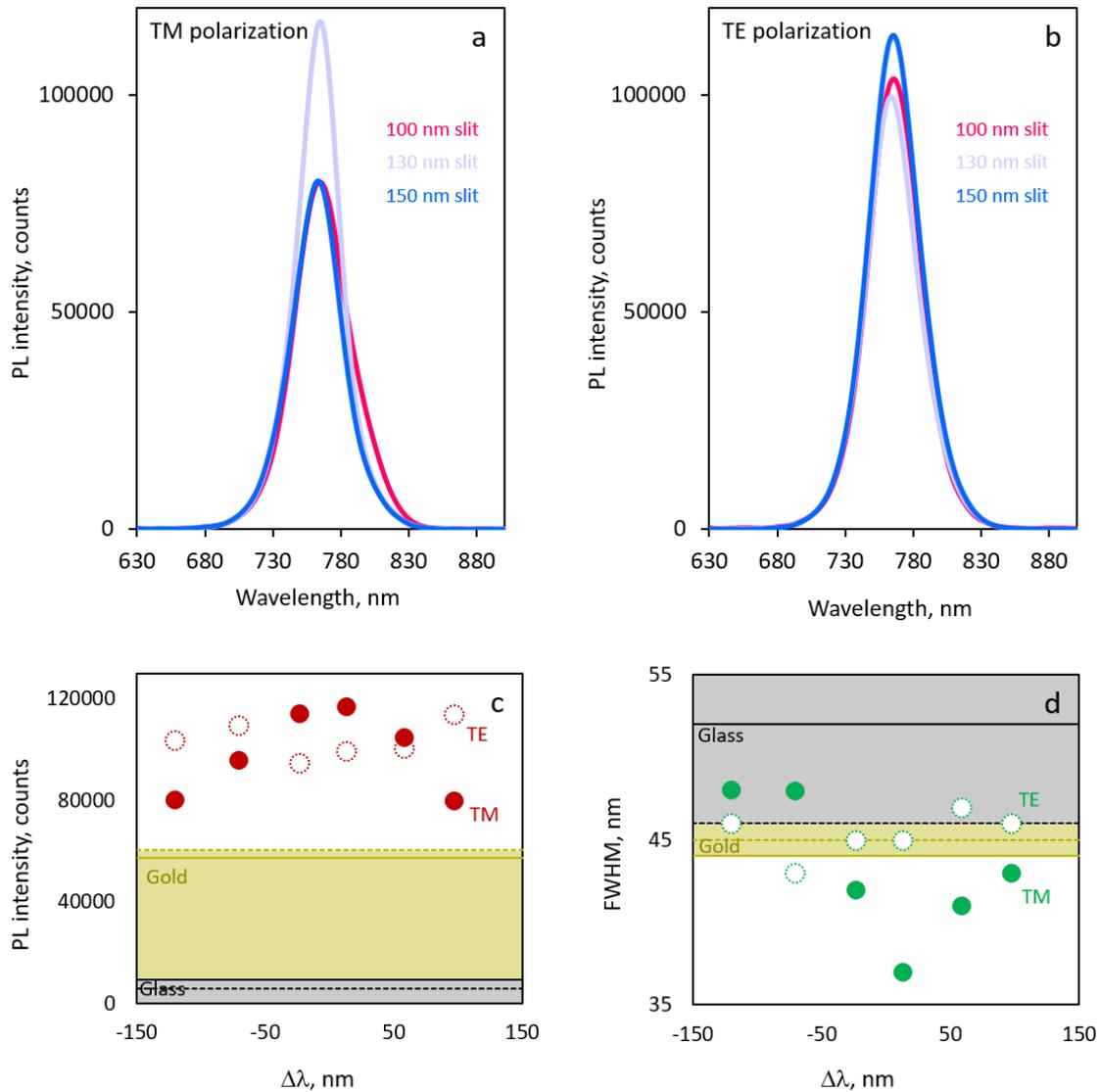

**Figure 3**: TM (a) and (b) TE (b) polarized spectra of a 65nm thick MAPbI$_3$ perovskite film on metamaterials with slit lengths of 100 nm (magenta), 130 nm (lilac) and 150 nm (blue); MAPbI$_3$ intensity (c) and linewidth (d) dependence as function of the wavelength mismatch between the metamaterial optical absorption resonance and the perovskite emission peak for both TM (full markers) and TE (dashed markers) polarizations [as reference, the horizontal lines and shaded areas indicate the intensities in TM (full lines) and TE (dashed lines) polarizations for the MAPbI$_3$ film on glass (black), on gold (yellow)].

Figures 3a and 3b show the photoluminescence intensity of the perovskite film on metamaterials, for arrays whose resonance is blue-shifted (100 nm slit, magenta curve), centred (130 nm slit, lilac curve), and red-shifted (150 nm slit, blue curve) with respect to the MAPbI$_3$ emission peak, in TM

(perpendicular to slits) and TE (parallel to slits) polarizations, respectively. It is quite evident that, in TM polarization, the photoluminescence intensity is highest when the metamaterial resonance and the perovskite emission peaks match, while in TE polarization there is no clear dependence on the metamaterial resonance wavelength. To visualize how the photoluminescence intensity depends on the wavelength mismatch ($\Delta\lambda$) between the nanoslits resonance and the MAPbI$_3$ emission, we plot in Figure 3c the photoluminescence peak intensity for i) the MAPbI$_3$ film on glass (black horizontal lines), ii) the MAPbI$_3$ film on gold/SiO$_2$ (yellow horizontal lines) and iii) the MAPbI$_3$ film on all the six nanoslit metamaterial arrays (red circles), in both TM (full circles) and TE polarization (dashed circles). The following observations can be made: i) The MAPbI$_3$ photoluminescence intensity increases (~6x) when the film is placed in the vicinity of gold; ii) The MAPbI$_3$ photoluminescence intensity is further enhanced, by more than one order of magnitude (~12x), when the film is placed on the metamaterial slits; iii) TM polarized photoluminescence shows a clear dependence on $\Delta\lambda$, while TE polarized photoluminescence is unaffected by the mismatch between the metamaterial optical absorption resonance and the perovskite emission peak wavelengths. This indicates that resonant nanoslits cause a Purcell enhancement of the photoluminescence by confining the electrical field in a nanometric volume when they are excited with polarization perpendicular to the long axis of the slit[21]. Further evidence of the effect induced on the MAPbI$_3$ photoluminescence by the photonic mode's volume confinement is provided by the linewidth of the TM polarized photoluminescence emission, which decreases proportionally to $\Delta\lambda$, as shown in Figure 3d. The fact that photoluminescence enhancement is comparable for TM and TE polarizations shall be attributed to the non-conformal coverage of the nanoslit volume by the perovskite films, as explained in Figure S1 (Supplementary Information). This also suggests that the values reported here are only a lower bound for achievable enhancement factors.

A direct manifestation of the enhancement of radiative emission rates by the Purcell effect is the reduction of photoluminescence lifetimes.[28,29] The photoluminescence decay traces for the MAPbI$_3$ perovskite film on metamaterial arrays with nanoslits of increasing length are plotted in Figure

4a, together with the fitting curves overlaid to the corresponding spectra (the spectra are vertically shifted for ease of visualization). The peak intensities of the decay traces show a clear dependence on Δλ (Figure 4b), in good agreement with the behaviour of steady-state spectra. The photoluminescence of MAPbI$_3$ films on glass shows a double-exponential decay with characteristic lifetimes of $\tau_1 \approx 0.29$ ns and $\tau_2 \approx 4.8$ ns, which are in good agreement with previous assignments to Auger recombination ($\tau_1$) and charge carrier relaxation through band-edge emission ($\tau_2$)[1].

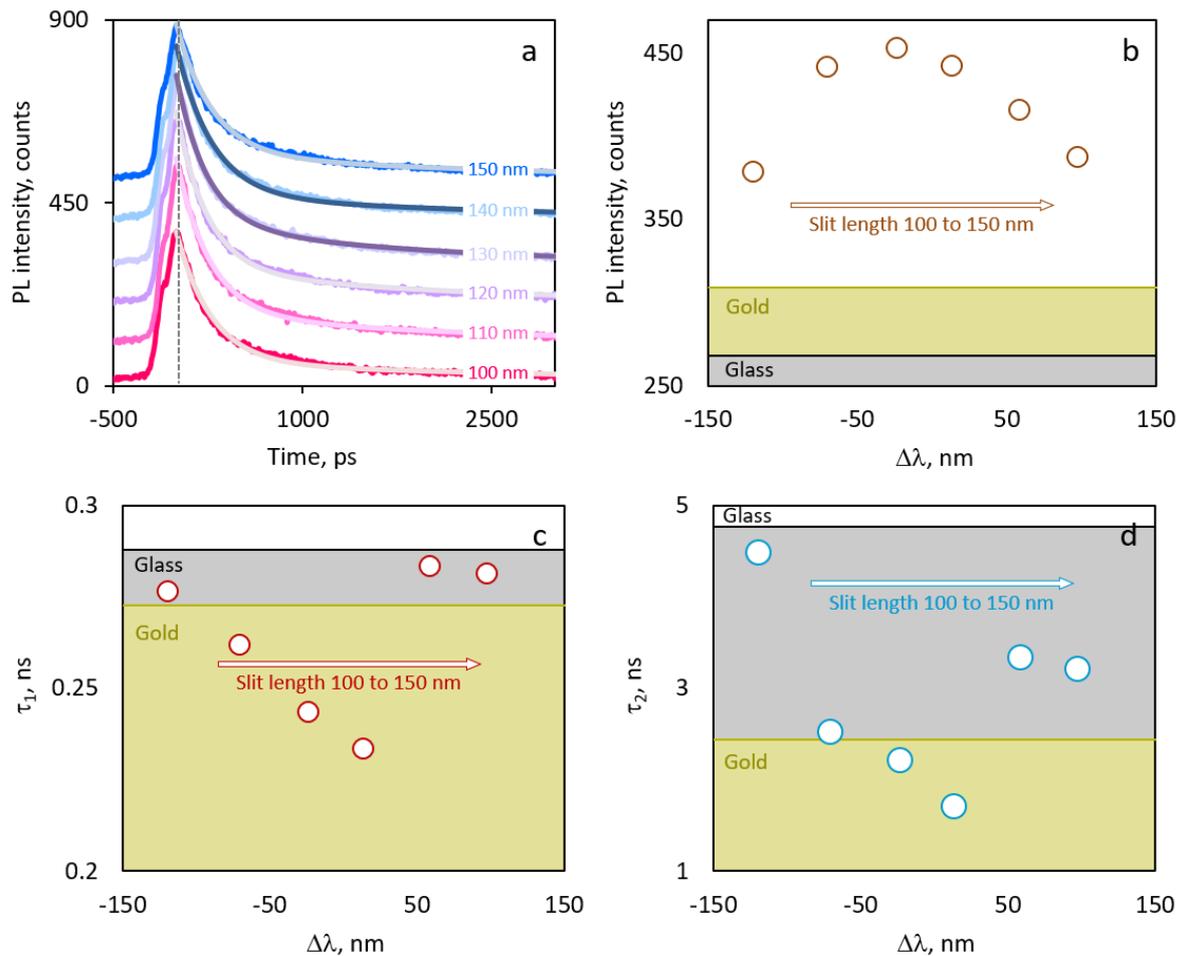

**Figure 4**: a) Time resolved photoluminescence of a 65nm thick MAPbI$_3$ perovskite film on metamaterials with slit lengths varying from 100 nm to 160nm, and fitting curves overlaid, (spectra are vertically shifted for ease of visualization); Time resolved photoluminescence peak intensity (b) and 2-exponential fitting time dependence (c & d) of MAPbI$_3$ on metamaterials as function of the detuning of the metamaterial resonance from the perovskite emission peak [as reference, the horizontal lines and shaded areas indicate the PL intensities for the MAPbI$_3$ film on glass (black), on gold (yellow)].

The hybridization of the MAPbI$_3$ film with the nanoslit metamaterials induces a significant reduction of both $\tau_1$ and $\tau_2$ (Figures 4c and 4d, respectively), consistent with the dependence of photoluminescence intensity and linewidth on $\Delta\lambda$. The three-fold shortening of radiative lifetime $\tau_2$ achieved at maximum overlap ($\Delta\lambda\sim0$, when the metamaterial resonance wavelength matches the MAPbI$_3$ emission peak) corresponds to a Purcell factor of nearly 3. We also observe a slight reduction (~20%) of the Auger recombination lifetimes, $\tau_1$, correlated with the metamaterials resonances, which may be attributed to plasmon-induced hot carrier generation[21]; given the more significant reduction in $\tau_2$, employing of metamaterials is clearly advantageous to enhance photoluminescence emission. Note that the Purcell factor reported here shall be considered as a lower bound, since time-resolved data were acquired for unpolarised emission while steady-state measurements indicate that nanoslit cavities affect almost exclusively TM polarized photoluminescence. Furthermore, all measurements were performed at room temperature, and a further increase in radiative rates could be expected at lower temperatures. Thus, the improvements in luminescence properties of perovskite films coupled to resonant metamaterial structures registered so far (enhancement of photoluminescence intensity, narrowing of steady-state linewidths and shortening of SE lifetimes) could arguably lead to a reduction of amplified spontaneous emission (ASE) and lasing thresholds, an important step towards integrated perovskites light emitting devices.

In summary, inspired by the proven ability of metamaterials to manipulate radiation, promote light-matter interaction and facilitate emission at any wavelength and across numerous types of media, we hybridized a metal-halide perovskite film with nanoslits gold metamaterials to control and enhance its optical emission. We demonstrated a substantial increase of the intrinsic photoluminescence intensity by more than one order of magnitude when the metamaterial resonance was tuned to the perovskite intrinsic emission peak. This increase in photoluminescence intensity is accompanied by a significant reduction in the spontaneous emission lifetimes, clearly indicating that the enhancement is radiative in nature. Improved performance and additional functionalities can be foreseen by optimization of the metamaterial structures and their coupling to the perovskite films (e.g., better permeation of the perovskite precursors into the nanostructured scaffold before conversion and crystallization). Similar design principles could also be adopted in the layout of electrodes[30,31]. Thus, our strategy hints potential reductions of amplified spontaneous emission (ASE) and lasing thresholds, paving the way towards on-chip integrated perovskite photonic components with wide spectral tunability from UV to NIR, and could potentially constitute a viable architecture for perovskite-based integrated light emitting devices, both optically and electrically pumped.

**Methods**

**Device Fabrication.** The nanoslits metamaterials were fabricated by depositing a 30 nm thin film of gold, via thermal evaporation, over a glass coverslip substrate, followed by focused ion beam (FIB) milling, with a FEI Helios 650 dual FIB/SEM system, of the metamaterial arrays of nanoslits of increasing length and the constant width (nominally ~30 nm), and a second round of thermal evaporation of ~5 nm silica. The substrates were then exposed under UV ozone cleaning for 20 minutes before perovskite deposition. MAPbI$_3$ perovskite film was deposited by spin coating from 0.33M solutions. In a typical example, PbI$_2$ (89.12 mg, TCI, 99.99%), methylammonium iodide MAI (30.73 mg, Dyesol) and 13.7 μl of dimethyl sulfoxide DMSO (anhydrous, Sigma Aldrich) were dissolved in 580 μl of dimethyl formamide DMF (anhydrous, Sigma Aldrich). The solution was heated at 100 °C for 1h, and then spin-coated on the substrate at 4000 rpm, 15 s. Solvent quenching with 300 μl of toluene (anhydrous, Sigma Aldrich) was performed after 6s of rotation. The films were then annealed for 15 min at 100 °C on a hotplate. All the perovskite deposition procedure was performed in a glove-box under Ar inert atmosphere.

**Optical Measurements.** The normal-incidence absorption spectra of the metamaterials were measured, for incident polarizations perpendicular and parallel to the main axis of the nanoslits (TM and TE orientations, respectively), using a microspectrophotometer (Jasco MV2000), through a 36× objective with a circular sampling aperture size of 25 μm × 25 μm. Steady-state photoluminescence spectra were measured under a microscope set-up and flow of nitrogen to avoid degradation of the MAPbI film due to moisture absorption. The samples were pumped using a PicoHarp 405 nm diode ps laser, with repetition rate of 40 MHz, focused down to a spot of ~20 μm diameter. The photoluminescence emission was polarized and detected by an ACTON spectrometer with an integration time of 100 ms per wavelength.

**Time resolved photoluminescence.** Time resolved PL measurements were done by exciting samples with 400nm wavelength, 100 fs pulse duration and 1000 Hz repetition rate femtosecond laser pulses. Femtosecond laser pulses were generated by Spectra Physics amplified laser system (MaiTai, Spitfire Ace). SHG crystal was used to generate 400nm wavelength. PL signals from perovskite film on metamaterial samples were collected and collimated before focusing in to the Optronics streak camera (SC 10). Temporal resolution of this streak camera is ~50 ps.

**Simulations.** Optical spectra and field maps were generated by full-wave electromagnetic simulations using COMSOL Multiphysics. Literature data were used to describe both the silica substrate, the gold film and the 5 nm silica overcoating, while experimental ellipsometric values were used for the perovskite film[6]. The samples were described as infinitely extended by using periodic boundary conditions and illuminated at normal incidence. The emission spectra were simulated by placing an array of infinitesimally small dipoles within the perovskite film, with spectral distribution mimicking the steady-state photoluminescence spectrum of the flat perovskite film. The total emission power was determined by averaging individual dipole contributions at each wavelength, weighted by the simulated field intensity generated at the dipole position by the pump beam ($\lambda$ = 405 nm) to account for the absorption[8,25].

**Acknowledgements**

This work was supported by the Singapore Ministry of Education Tier 3 grant (MOE2016-T3-1-006), the Singapore National Research Foundation, Prime Minister's Office, under its Competitive Research Programme (CRP Award No. NRF-CRP14-2014-03), and by the A*STAR-AME programmatic grant on Nanoantenna Spatial Light Modulators for Next-Gen Display Technologies (Grant No. A18A7b0058).

**Supplementary Information**

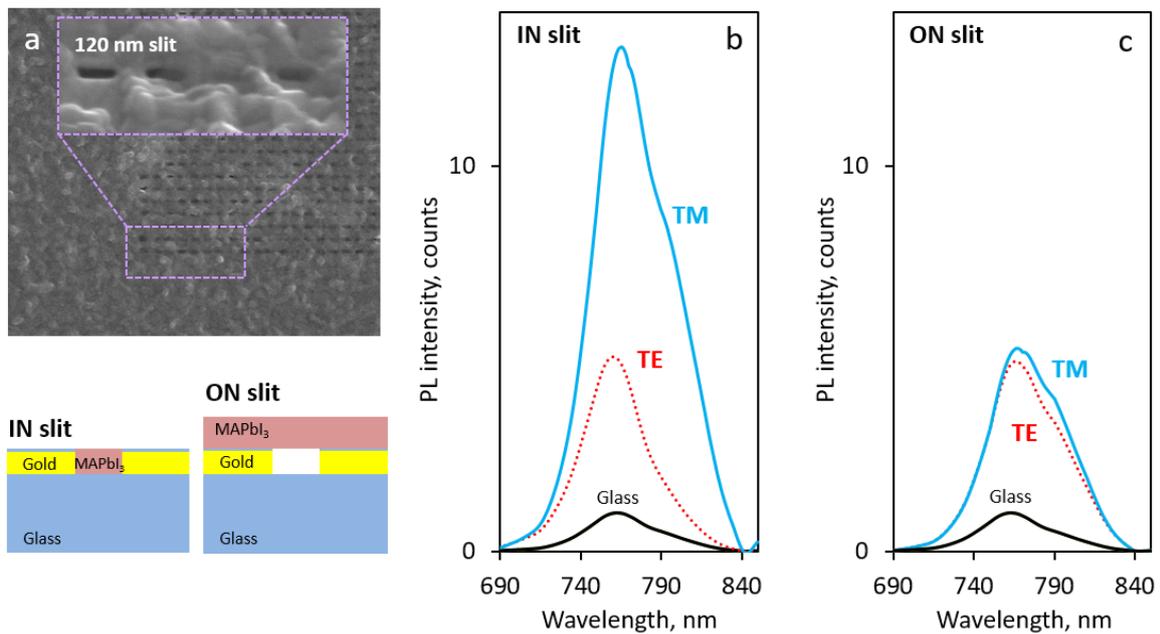

**Figure S1**: Secondary electrons image (a) of 120nm slit metamaterials carved by FIB milling on a 30nm Au + 5nm SiO$_2$ film, on glass substrate, and coated with 65nm thick MAPbI$_3$ perovskite film: the perovskite film appears to be partially IN the slit and partially ON the slit, as exemplified by the schematics; and full wave COMSOL simulations of photoluminescence enhancement for the perovskite film IN (b) and ON (c) the metamaterial slit, for both TM (blue full curves) and TE (red dotted curves) polarizations, with respect to the emission of perovskite film on glass.